\documentclass[journal]{IEEEtran}
\usepackage{hyperref}
\usepackage{graphicx}
\usepackage{amsmath}
\usepackage{subcaption}
\usepackage{changepage} % For adjustwidth environment
\usepackage{multirow}

\usepackage{arydshln}

\ifCLASSINFOpdf
\else
\fi
\graphicspath{ {Images/} }

\begin{document}
\title{Efficient Feature Extraction Using Light-Weight CNN Attention-Based Deep Learning Architectures for Ultrasound Fetal Plane Classification}

\author{Arrun Sivasubramanian\IEEEauthorrefmark{1}, Divya Sasidharan\IEEEauthorrefmark{1}, Sowmya V\IEEEauthorrefmark{1}, Vinayakumar Ravi\IEEEauthorrefmark{2}% <-this % stops a space

\IEEEauthorblockA{\IEEEauthorrefmark{1}Amrita School of Artificial Intelligence, Coimbatore, Amrita Vishwa Vidyapeetham, India.\\
\IEEEauthorrefmark{2}Center for Artificial Intelligence, Prince Mohammed Bin Fahd University, Khobar, Saudi Arabia\\
}

\thanks {Email: Arrun Sivasubramanian (arrun.sivasubramanian@gmail.com), Divya Sasidharan (s\_divya1@cb.amrita.edu), Sowmya V (v\_sowmya@cb.amrita.edu) and Vinayakumar Ravi (vravi@pmu.edu.sa)}}

\maketitle

\begin{abstract}

Ultrasound fetal imaging is beneficial to support prenatal development because it is affordable and non-intrusive. Nevertheless, fetal plane classification (FPC) remains challenging and time-consuming for obstetricians since it depends on nuanced clinical aspects, which increases the difficulty in identifying relevant features of the fetal anatomy. Thus, to assist with its accurate feature extraction, a lightweight artificial intelligence architecture leveraging convolutional neural networks and attention mechanisms is proposed to classify the largest benchmark ultrasound dataset. The approach fine-tunes from lightweight EfficientNet feature extraction backbones pre-trained on the ImageNet1k. to classify key fetal planes such as the brain, femur, thorax, cervix, and abdomen. Our methodology incorporates the attention mechanism to refine features and 3-layer perceptrons for classification, achieving superior performance with the highest Top-1 accuracy of 96.25\%, Top-2 accuracy of 99.80\% and F1-Score of 0.9576. Importantly, the model has 40x fewer trainable parameters than existing benchmark ensemble or transformer pipelines, facilitating easy deployment on edge devices to help clinical practitioners with real-time FPC. The findings are also interpreted using GradCAM to carry out clinical correlation to aid doctors with diagnostics and improve treatment plans for expectant mothers.% For ease of reproducibility, the codes of the work will be made available in the GitHub link: \href{}{https://github.com/argon125/US-Fetal-Plane-Classification-Using-Light-Weight-DL-Architectures}.

\end{abstract}

\begin{IEEEkeywords}
Ultrasound, Fetal Plane Classification, EfficientNet, Attention, GradCAM, SHAP.  
\end{IEEEkeywords}

% \fi
\IEEEpeerreviewmaketitle

\section{Introduction}\label{sec1}
\IEEEPARstart{U}{ltrasound} (US) is a non-intrusive technique that employs sound waves to produce imagery of the inner anatomy. It is particularly well-known for tracking a fetus's progress throughout pregnancy \cite{ref1}. Fetal development is often tracked using several metrics derived from maternal-fetal scans during regular clinical obstetric exams \cite{ref5} and identifying anomalies to give obstetricians critical information to guarantee the best possible outcomes for the mother and the unborn child  \cite{ref6}. However, over the past 20 years, though the rate of maternal death has decreased globally, there is a significant gap between high and low-income nations due to a shortage of well-trained gynaecologists and less frequent pregnancy diagnoses in developing nations, leading to maternal fatalities accounting for almost 94\% of all deaths \cite{ref4}. Furthermore, because ultrasound pictures have poor contrast, low imaging quality, and a high degree of variability, it can be challenging for medical practitioners to understand every aspect of the images \cite{ref49}. 

Most nations provide at least one mid-trimester ultrasound scan as part of routine prenatal care to detect any potential pregnancy-related problems \cite{ref7}, with fetal biometry, fetal weight estimation, and doppler blood flow being the biomarkers most often utilized by clinicians to detect fetal anomalies \cite{ref2}. However, this process is arduous and error-prone because each screening ultrasound examination typically consists of over 20 images and even expert sonographers find it challenging to obtain standard fetal ultrasound planes with accurate fetal anatomical structures \cite{ref3}. Furthermore, if the fetal position is not ideal, it may be challenging to get a high-quality image of the intended anatomical structure \cite{ref19}. These challenges highlight the crucial requirement for interventions to decrease maternal mortality rates, and an automated system doing this work would be more economical and less prone to mistakes. This intelligent system will support clinical decision-making, and healthcare workers may analyze medical images to find anomalies quickly. 

Artificial intelligence (AI) has become a potent instrument for improving many industries, including manufacturing, transportation, healthcare, and finance \cite{ref30}. AI has the capacity to evaluate patient data and carry out activities that generally require human intellect, such as pattern recognition, language comprehension, and problem-solving, and holds great promise in performing precise diagnoses by improving patient diagnosis and treatment by assisting medical personnel in improving patient outcomes \cite{ref50}. Deep learning (DL) and CNNs have gained popularity recently as popular solutions for various medical image classification issues, including FPC and other diseases such as Alzheimer's \cite{ref8}, lung nodules in CT/X-ray \cite{ref9}, skin and breast lesions \cite{ref10,ref11}, echo-cardiogram views \cite{ref12}, to name a few. 

Although current models for FPC show promising metrics, there is always room for improvement due to the diverse and challenging nature of ultrasound images. Attention mechanisms can enhance DL models by allowing them to focus on specific parts of the input data, thereby improving accuracy and efficiency \cite{ref51}.  The increased complexity and variability in these images necessitate more advanced techniques for feature extraction, such as leveraging a CNN-Attention pipeline for this task. To the best of our knowledge, no such mechanism in the literature has performed this for the FPC task. Also, the DL models proposed so far contain a lot of trainable parameters, and their size hinders deployment in mobile devices, rendering them useless for practical applications. Thus, the authors of this work propose a lightweight architecture to assist doctors with their diagnosis. Moreover, if there is a difference in opinion among the experts, the decisions made by a black-box AI algorithm should be explainable, which has been addressed by us with an explainable AI framework. This will help the experts arrive at a consensus with automated solutions in record time.

The major contributions of this work are: 

\begin{itemize}
  
  \item Engineer a CNN-Attention pipeline using feature extraction backbones from the EfficientNet family and different attention mechanisms to refine the output feature maps before classification with three multi-layer perceptrons. The pipeline contains 40x lesser trainable parameters to ensure early diagnosis and easy deployability on small storage-size or portable medical devices. 

  \item Using transfer learning from state-of-the-art models pre-trained on the ImageNet1k dataset to fine-tune the model. This results in an improvement in the performance of automated models to perform fetal plane classification on the most extensive ultrasound FPC dataset, which contains 12,000 images.

  \item A comparison in terms of quantitative metrics such as accuracy, precision, recall, and F1-score between the suggested approach and the current benchmark models for FPC automation in the literature. In this study, additional insights into data with metrics like confusion matrix and ROC-AUC curves are obtained to understand the efficacy of the model on every fetal plane category of the dataset.
  
  \item Including the interpretability of these results using GradCAM heatmaps. The results from this explainable AI framework would assist obstetricians/specialists who inspect maternal ultrasounds in accurately determining fetal abnormalities to determine the correct plan of medical interventions. It also helps determine why the proposed model classified a particular ultrasound image into the corresponding fetal plane. 
\end{itemize}

The manuscript is structured as follows: Section 2\ref{sec2} elucidates the relevant work utilizing several datasets and their methodologies for automating the FPC task. Section 3\ref{sec3} covers the description of the various CNN feature extraction backbones from the EfficientNetBx and EfficientNetV2x family, the attention mechanisms leveraged, and the methodology adopted in this work. In Section 4\ref{sec4}, the results are elaborated upon using a range of quantitative and qualitative criteria, and the advantages and limitations of this study are discussed. In the final Section 5\ref{sec5}, we conclude the work and provide potential directions for further research.

\section{Related Works} \label{sec2}

Performing FPC using AI algorithms is important because it can enhance the efficiency and accuracy of prenatal ultrasound examinations by overcoming limitations. Additionally, it can contribute to the accurate and early detection of abnormalities, allowing for timely intervention and better management of pregnancy complications. This aids healthcare professionals in obtaining standardized and consistent imaging, improving the overall quality of prenatal care. The works done in the literature to perform fetal plane classification can be classified into machine learning (ML) and deep learning (DL) techniques. 

A few core ML methods employed to perform FPC do find an important place even in the present era. Six fetal ultrasound planes were classified using a random-forest-based technique by Yaqub et al. \cite{ref21}. Carneiro et al. \cite{ref22} employed a probabilistic boosting tree to identify and categorize fetal morphological features in a related study. Yaqub et al. \cite{ref23} also created a method in a different study that examines if an individual's imaging record has all pertinent anatomical prenatal images. Spatial-temporal techniques are used by Yasrab et al. \cite{ref24} and Chen et al. \cite{ref25} to categorize fetal ultrasound images at the frame level. However, a lot of emphasis in recent times has been placed on deep learning-based techniques because of their automatic ability to retrieve features without physical kernels designed by humans, the diversity of data, and the easily available computing power.

There are several deep-learning feature extraction networks used for FPC. Different approaches to the localization of fetal ultrasound images have been published based on variations of VGG feature extraction backbones. The authors proposed a unique CNN architecture known as SonoNet in \cite{ref15}. VGG-net \cite{ref16} architecture serves as the foundation for SonoNet's design. The 13 fetal standard views are classified and localized using a bounding box by SonoNet and its variations, utilizing 2-D and 3-D fetal ultrasound picture data. In every SonoNet variation, convolutional layers take the role of the last fully connected layers to minimize the number of parameters. The transfer learning technique has been used in \cite{ref17} to construct an automated framework for fetus face standard plane recognition utilizing pre-trained VGG. Additionally, a fine-tuning process analysis was conducted on VGG-16 and VGG-19. The last max-pooling layer has replaced the global average pooling (GAP) layer. To lessen the problems with overfitting, the batch normalization layer was positioned before the layer with the ReLU layer, which introduces non-linearity to the data. VGG-19-GAP is the name given to the altered VGG architecture. Based on the findings, the VGG-19-GAP performed better. An approach for computer-aided diagnostics has been given in \cite{ref18} to identify five frequent anomalies in the embryonic brain. Fetal scans segment the craniocerebral areas, which are divided into four groups. Moreover, class activation maps were used to locate the aberrant photos. 

Additionally, studies utilizing smaller, independent datasets and other popular feature extraction backbones have been conducted to categorize fetal planes. Using a testing set of 5678 ultrasound pictures, Kong et al. \cite{ref26} employed multi-scale dense networks with classification metrics as high as 98\% to identify four different fetal anatomies. Using a network known as SPRNet, Liang et al. \cite{ref27} employed an automated technique for identifying and diagnosing fetal planes. Based on DenseNet, the network was trained using data-based partial transfer learning on pictures from the placenta and fetal ultrasounds. According to the findings, SPRNet obtains accuracy, recall, specification, and F1 of 0.99, 0.96, 0.99, and 0.95. In order to improve fetal brain classification using ResNet, Montero et al. \cite{ref28} used generative adversarial networks (GANs). ResNet produced an AUC of 0.86, Accuracy, and F1 of 0.81 and 0.80, respectively. Meng et al. \cite{ref29} utilized improved feature alignment to extract discriminative and domain-invariant features across domains to conduct a cross-device classification of planes. Certain research concentrates on identifying a single fetal standard plan. There's a chance this won't apply to all ultrasound scan kinds. Other research has investigated the application of ultrasound in the identification of fetal planes in biometry-related clinical situations, including the determination of fetal gestational age \cite{ref32,ref33} and the identification of fetal gender \cite{ref31}.

Given that deep features extracted from an ensemble of various CNNs improve the feature representation, an automated classification technique with deep feature integration extracted using AlexNet and VGG-19 modules is presented in \cite{ref34}. A multi-layer perceptron to classify fetal ultrasound images. Although the proposed method achieves satisfactory results, it has some limitations.  With the exception of the maternal cervix and fetal brain, the three other classes—the abdomen, femur, and thorax—are occasionally misdiagnosed as belonging to different classes. However, implementing transfer learning of features \cite{ref39} from prominent classification datasets such as ImageNet \cite{ref40} and implementing picture cropping or pre-processing might help to enhance the inter-class variance. Nevertheless, geometric augmentations have helped in improving metrics \cite{ref34}. Additionally, the suggested method's classification performance may be improved by employing dimensionality reduction techniques to reduce the integrated feature size and make a lightweight architecture. 

As transformers, a CNN substitute that, via its self-attention method, can identify salient feature representations and extract long-range relationships from medical images \cite{ref35}, the work done in \cite{ref36} is based on a transformer-based method that uses a newly created residual cross-variance attention block to utilize global and spatial information for FPC. This block presents an enhanced cross-covariance attention (XCA) technique to extract a long-range representation from the input utilizing spatial and global information. Very dark and blurred images, which correspond to a dull anatomical region in the ultrasound image dataset, can lead to incorrect predictions; the training process proposed neglects a total of 615 images out of the unaugmented 12400 input images with a brightness score lower than 20 and blurriness score of less than 100. Post-data augmentation techniques were used to normalize all the data based on the average and standard deviation of pixel intensity for the ultrasound image to avoid overfitting. The experimental findings demonstrate that COMFormer achieves 95.64\% classification accuracy on fetal anatomy, outperforming the latest CNN and transformer-based models. Yet, as shown in \cite{ref37}, deep feature integration using a stacked ensemble of DL models with three pre-trained feature extraction backbones—AlexNet, VGG-19, and DarkNet-19—shows that the suggested method performed well, achieving 94.02\% accuracy, 96.28\% recall, and 95.08\% F1-score. The hard voting method is used to determine the final predictions after extracting the classification scores of different backbones using random forest, a bagging ML classifier. However, the approach is computationally expensive, with a lot of trainable parameters, as it leverages four heavy-weight backbones of the three architectures to perform feature extraction before an ensemble technique to perform voting-based classification. 

In summary, a number of studies have shown that ultrasonography may be used to classify maternal fetal standard plans, showing promising outcomes for both 2D and 3D methods. The existing works in the literature mostly focus on enhancing classification metrics by using either pre-trained CNNs or transformers. However, no proposed mechanism for FPC combines the CNN-Attention paradigm to extract relevant features from complex ultrasound images. Moreover, these models are computationally complex and contain a lot of trainable parameters to improve metrics. Even with studies that use extensive usage of individually trained deep CNN models (e.g., VGG-Nets and AlexNet) through transfer learning, the authors think that detection accuracy may still be improved with lighter backbones. Most of these studies also do not help in interpreting the results using explainable AI frameworks to perform clinical corellations. Thus, the difficulties and constraints that still need to be overcome for the FPC task can be summed up as follows.

\begin{itemize}
    \item A small number of standard and non-standard fetal planes have structural similarities, yet it is crucial to build a robust yet lightweight automated classifier using deep learning to overcome the challenges present in ultrasound imaging and FPC.
    
    \item The correlation amongst fetal ultrasound images makes it difficult, even for experienced sonographers, to identify and select the standard fetal view during prenatal screening. Thus, rapidly extracting key deep features and identifying the relevant feature maps from them with additional mechanisms could help in efficient classification.

    \item Building an automated classifier that can demystify the black box nature of models that accurately predict fetal plane to provide insights into maternal and fetal health prognoses.
\end{itemize}

In response to identified drawbacks in existing literature within our proposed domain, our research systematically addresses these concerns:

\begin{itemize}
    \item A computationally lightweight backbone to perform feature extraction that will reduce the training time without compromising on accuracy is quintessential to performing accurate and faster real-time fetal plane classification.
    
    \item To the best of our knowledge, no mechanism explores the paradigm of utilizing the attention mechanism as used in transformers, combined with the prowess of feature extraction using convolution in CNNs. This is leveraged post-feature extraction from a lightweight pre-trained CNN backbone on which transfer learning is performed. The multi-layered perceptron (MLP) refines the predictions even further, as justified in the results.
    
    \item Explainability of the features using GradCam heatmaps have not been performed in many works, and it would assist the doctors in understanding what that particular fetal biometry was classified into a particular plane and support clinical correlations related to prenatal health. 
\end{itemize}

Through comprehensive analysis and experimentation, we provide novel solutions and insights that effectively mitigate or overcome the previously identified limitations. Our findings not only contribute to advancing knowledge within the field but also serve to address the gaps highlighted by the existing literature.

\section{Methodology}\label{sec3}

\subsection{Dataset Description}\label{sec3.1}

\begin{figure}[h]
    \centering
    \includegraphics[width=\linewidth]{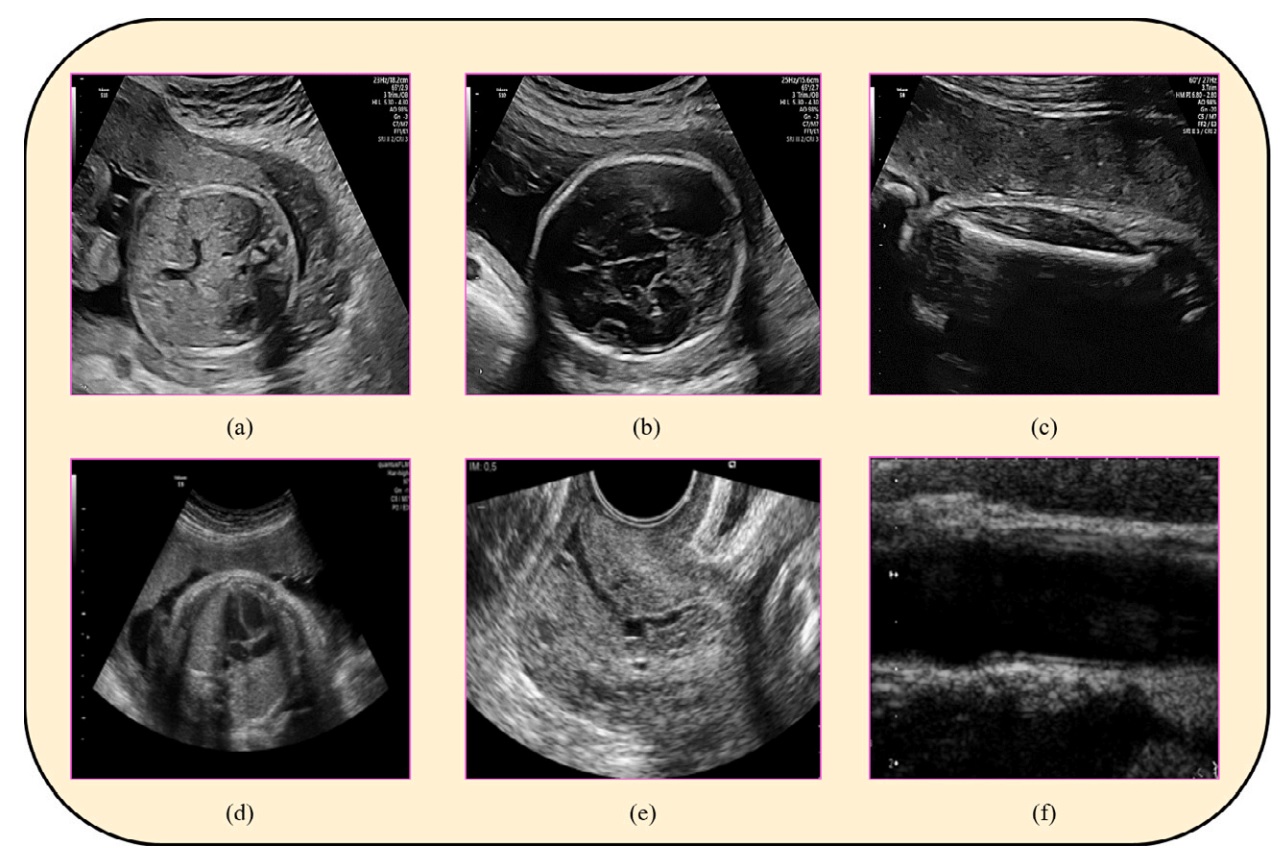}
    \caption{Sample input ultrasound fetal places belonging to all six classes: (a) Fetal Abdomen, (b) Fetal Brain, (c) Fetal Femur, (d) Fetal Thorax, (e) Maternal Cervix, and (f) Others) of the dataset.}
    \label{fig1}
\end{figure}

The BCNatal hospital in Spain, which has sizable specialized maternal-fetal departments that handle thousands of births annually, provided the dataset utilized for this study \cite{ref42}. The research encompasses all expectant mothers who attended routine prenatal appointments in the second and third trimesters. All the data was claimed to be collected by following all due protocols established by the rules and after obtaining each patient's permission for research purposes. The sample input images for each fetal place present in the dataset are shown in Figure \ref{fig1}. It also contains information about the number of samples after the 80:20 train-test split that was performed to train and validate the models of the work.

The operators were to refrain from utilizing any post-processing technique, including noise smoothing. They were also free to choose how to adjust the remaining picture settings, which included gain, frequency, and gain compensation. Finally, 12400 fetal ultrasounds were collected and annotated into six different classes. This is the largest publicly available and sizeable dataset that can be used to automate the task of FPC, as most other datasets are smaller with less diversity of images or are private. Table \ref{tab1} shows the distribution of data among the maternal-fetal parts captured for the ultrasound dataset.

\begin{table}[h]

    \centering
    \caption{Quantitiative description of the fetal US dataset curated by BCNatal.}
    \label{tab1}
    \fontsize{8pt}{10pt}\selectfont
    \begin{tabular}{|c|c|c|c|c|}
    \hline
    \textbf{Anatomical} & \textbf{Num. of } & \textbf{Num. of } & \textbf{Train} & \textbf{Test} \\ 
    \textbf{ Plane}  & \textbf{ Patients} & \textbf{Planes} & \textbf{ Samples} & \textbf{Samples}\\
    \hline
    Fetal Abdomen (FA) & 595 & 711 & 569 & 142\\
    Fetal Brain (FB) & 1,082 & 3,092 & 2474 & 618\\
    Fetal Femur (FF) & 754 & 1,040 & 832 & 208\\
    Fetal Thorax (FT) & 755 & 1,718 & 1374 & 344 \\
    Maternal Cervix (MC) & 917 & 1,626 & 1301 & 325 \\
    Others (O) & 734 & 4213 & 3370 & 843 \\
    \textbf{Total} & \textbf{1792} & \textbf{12400} & \textbf{9920} & \textbf{2480}\\
    \hline
    \end{tabular}
\end{table}

\begin{figure*}[h]
    \centering
    \includegraphics[width=\textwidth]{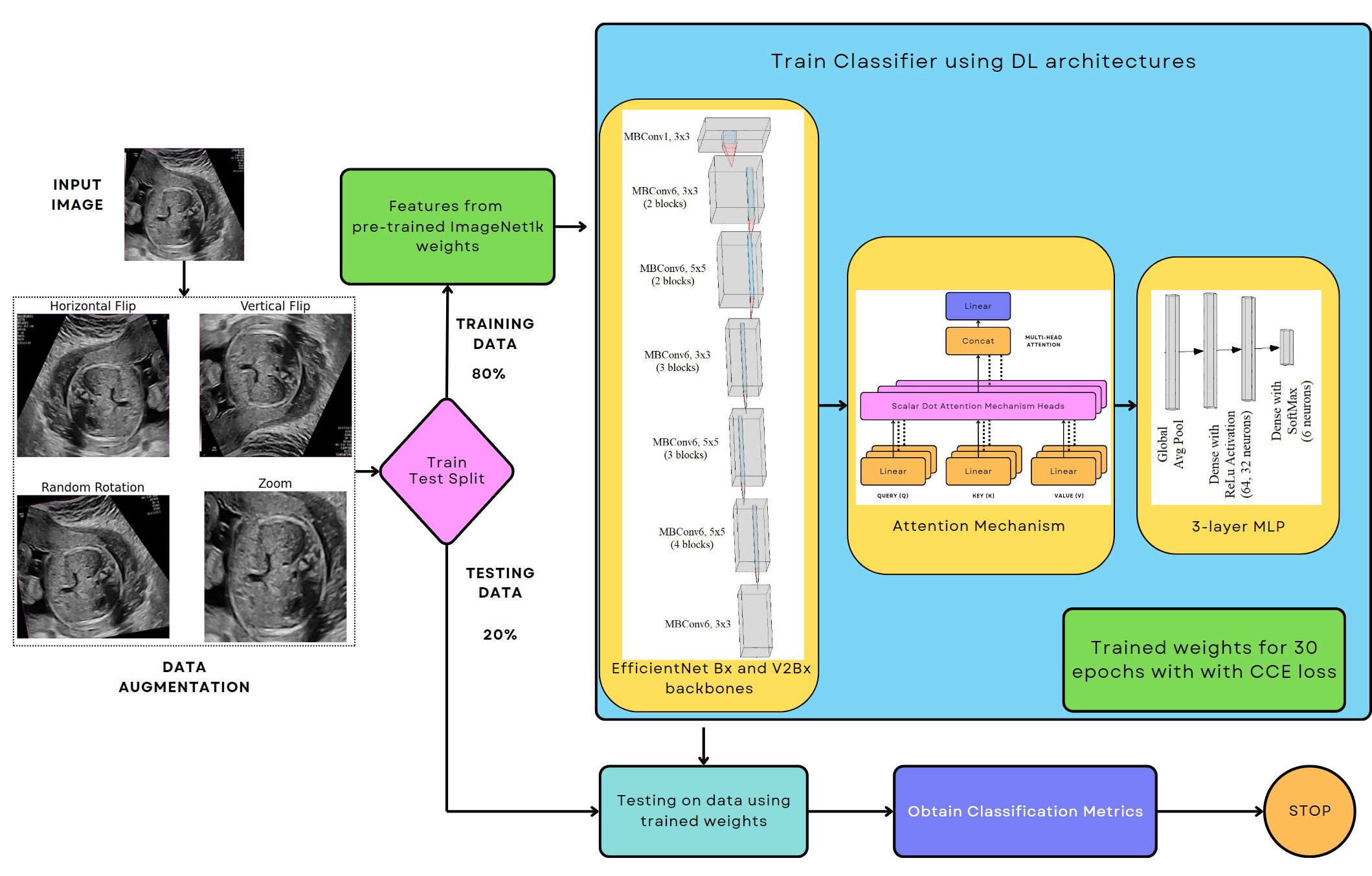}
    \caption{Pipeline of the proposed lightweight architecture for the US FPC task.}
    \label{fig2}
\end{figure*}

\subsection{Deep learning feature extraction backbones}\label{sec3.2}

DL models are becoming increasingly popular due to their ability to automatically extract relevant features from the input data for several tasks, including biomedical image analysis. The EfficientNet is a family of CNN architectures for feature extraction that is designed to achieve state-of-the-art performance due to its compound scaling method, which balances network depth, width, and resolution for optimal performance with significantly fewer parameters compared to traditional models \cite{ref41}. The variants introduced in the first cycle contained eight variations from B0 to B7, while the second version (V2) contains seven variations from V2B0 to V2B3 and Small, Medium, and Large, depending on the number of layers and trainable parameters. In summary, the integration of MBConv inverted bottleneck layers from MobileNet and Squeeze-and-Excitation (SE)  blocks to adaptively recalibrate channel-wise feature responses enables superior performance compared to other architectures in various applications. For this work, all backbones were trained and tested, while the explanation of the backbone yielding the benchmark result is given below:

\subsubsection{EfficientNetB0}\label{sec3.2.1}

EfficientNetB0 is the baseline model in the EfficientNet series. The network's depth, breadth, and resolution are consistently scaled with a compound scaling technique, which allows for better performance across various resource constraints. This makes EfficientNetB0 efficient for both inference and training as it employs a combination of depthwise separable convolutions and inverted residual blocks to extract features efficiently while maintaining computational effectiveness compared to other CNN counterparts. 

\subsubsection{EfficientNetV2B0}\label{sec3.2.2}

With upgrades to training methods and architectural design, EfficientNetV2B0 is an improved version of EfficientNetB0. It brings in new elements, including Swish activation functions and SE blocks, improving performance and feature representation. By adding a new expansion phase to its design, EfficientNetV2B0 extracts more relevant features from input images. This allows the network to convert inputs into a higher-dimensional space to extract more complicated features more effectively.

\subsection{Attention Mechanisms}\label{sec3.3}

Neural networks can be made to focus on particular portions of the input data by using an attention mechanism \cite{ref52}, which gives varied levels of priority to distinct features. In computer vision tasks, it is used to improve model performance by enabling it to attend to relevant patterns or pixels on an image selectively. In our study, it has been leveraged on the feature maps obtained after feature extraction from EfficientNet. The various types of attention used are described below:

\subsubsection{Scalar Dot Attention (SDA)}\label{sec3.3.1}

Scalar dot attention, sometimes called scaled dot-product attention, computes attention scores by scaling to avoid significant values after obtaining the dot product of the query and critical matrices. The attention output is subsequently produced by using these scores to weight the values matrix. The scaling factor ensures more stable gradients during training and better performance; it is usually the square root of the size of the critical matrices. The equation to compute scalar dot attention is shown in equation \ref{SDAeqn}.

\begin{equation}
\label{SDAeqn}
\text{SDA}(Q, K, V) = \text{Softmax} \left( \frac{QK^T}{\sqrt{d_k}} \right) V
\end{equation}

\subsubsection{Multi Head Attention (MHA)}\label{sec3.3.2}

Multihead attention is a method that helps pay attention to diverse sections of the input sequence at the same time. Unlike scalar dot attention, It computes distinct attention scores for each head after dividing the query, key, and value matrices into numerous heads. These scores are then concatenated and linearly transformed to obtain the final output. The equation to compute multi-head attention is shown in equation \ref{MHAeqn}. The value of each head in the equation is given by the scalar dot attention in \ref{SDAeqn}

\begin{equation}
\label{MHAeqn}
\text{MHA}(Q, K, V) = \text{Concat}(\text{head}_1, \text{head}_2, ..., \text{head}_h)W^O
\end{equation}

\subsubsection{Seq Self Attention (SSA)}\label{sec3.3.3}

Self-attention, or intra-attention, captures relationships within a sequence by allowing each element to weigh its relevance to every other element, including itself. This mechanism enables the model to evaluate the significance of each element in context, effectively capturing long-range dependencies and interactions. The components refine their representations based on their sequence positions, resulting in richer and more detailed embeddings. Self-attention is crucial for scenarios requiring an understanding of relationships across multiple sequence segments, theoretically outperforming other attention mechanisms used in similar tasks. The equation to compute SSA is shown in equation \ref{SSAeqn}.

\begin{equation}
\label{SSAeqn}
\text{SSA}(X) = \text{Softmax} \left ( \frac{XW_Q(XW_K)^T}{\sqrt{d_k}} \right)XW_V
\end{equation}

\subsection{XAI for explainability}\label{sec3.4}

\begin{table*}[h]
    \centering
    \caption{Comparison of the performance metrics of different benchmark architectures and the proposed model.}
    \label{tab3}
    \fontsize{8pt}{10pt}\selectfont
    \begin{tabular}{|c|c|c|ccccc|ccccc|}
        \hline
        \multirow{2}{*}{\textbf{Architecture}} & \multirow{2}{*}{\textbf{Parameters}} & \multirow{2}{*}{\textbf{Layers}} & \multicolumn{5}{c|}{\textbf{Without Attention}} & \multicolumn{5}{c|}{\textbf{With SeqSelfAttention}} \\ \cline{4-13} 
        & & & \textbf{T1-Acc} & \textbf{T2-Acc} & \textbf{Prec} & \textbf{Rec} & \textbf{F1} & \textbf{T1-Acc} & \textbf{T2-Acc} & \textbf{Prec} & \textbf{Rec} & \textbf{F1} \\ \hline
        EfficientNetB0 & \textbf{4,133,833} & \textbf{238} & 95.61 & 99.71 & 0.9440 & 0.9590 & 0.9511 & \textbf{96.05} & 99.60 & 0.9484 &  0.9586 & \textbf{0.9534} \\ 
        EfficientNetB1 & 6,659,501 & 340 & 95.49 & 99.60 & 0.9411 & 0.9566 & 0.9484 & 95.53 & 99.60 & 0.9430 &  0.9532 & 0.9476 \\ 
        EfficientNetB2 & 7,861,023 & 340 & 95.37 & 99.72 & 0.9376 & 0.9566 & 0.9466 & 95.45 & 99.60 & 0.9203 & 0.9575 & 0.9359 \\
        EfficientNetB3 & 10,884,181 & 385 & 95.41 & 99.64 & 0.9405 & 0.9545 & 0.9469 & 95.05 & 99.52 & 0.9382 & 0.9480 & 0.9427 \\ 
        EfficientNetV2B0 & \textbf{6,003,574} & 270 & \textbf{96.25} & \textbf{99.80} & \textbf{0.9529} & \textbf{0.9625} & \textbf{0.9576} & 95.33 & 99.64 & 0.9356 & 0.9622 & 0.9479 \\ 
        EfficientNetV2B1 & 7,015,386 & 334 & 95.81 & 99.68 & 0.9493 & 0.9550 & 0.9520 & 95.25 & 99.56 & 0.9333 & 0.9597 & 0.9456 \\ 
        EfficientNetV2B2 & 8,861,828 & 349 & 95.77 & 99.64 & 0.9421 & 0.9592 & 0.9501 & 95.81 & \textbf{99.64} & \textbf{0.9498} & 0.9543 & 0.9520 \\ 
        EfficientNetV2B3 & 13,031,397 & 409 & 95.65 & 99.60 & 0.9427 & 0.9574 & 0.9513 & 95.01 & 99.56 & 0.9420 & 0.9466 & 0.9441 \\ 
        EfficientNetV2S & 20,415,751 & 513 & 95.61 & 99.68 & 0.9367 & 0.9618 & 0.9483 & 94.89 & 99.60 & 0.9235 & \textbf{0.9627} & 0.9410 \\ 
        EfficientNetV2M & 53,234,779 & 740 & 94.28 & 99.45 & 0.9194 & 0.9486 & 0.9325 & 94.88 & 99.58 & 0.9150 & 0.9731 & 0.9407 \\ 
        EfficientNetV2L & 117,831,239 & 1,028 & 94.02 & 99.34 & 0.9219 & 0.9302 & 0.9223 & 94.14 & 99.45 & 0.9243 & 0.9587 & 0.9333 \\ \hline
    \end{tabular}
\end{table*}

\begin{table*}[h]
\centering
\caption{Comparison of metrics for different architectures across various attention mechanisms.}
\label{tab4}
\fontsize{8pt}{10pt}\selectfont
\begin{tabular}{|c|c|c|c|c|c|c|c|c|}
\hline
\multirow{2}{*}{\textbf{Architecture}} & \multicolumn{2}{c|}{\textbf{Metrics w/o attn.}} & \multicolumn{2}{c|}{\textbf{Metrics with SDA}} & \multicolumn{2}{c|}{\textbf{Metrics with MHA}} & \multicolumn{2}{c|}{\textbf{Metrics with SSA}} \\ \cline{2-9} & \textbf{T1-Acc} & \textbf{F1} & \textbf{T1-Acc} & \textbf{F1} & \textbf{T1-Acc} & \textbf{F1} & \textbf{T1-Acc} & \textbf{F1}\\
\hline
EfficientNetB0 & 95.61 & 0.9511 & 95.81 & 0.9513 & 95.93 & 0.9529 & \textbf{96.05} & \textbf{0.9534}\\
EfficientNetV2B0 & \textbf{96.25} & \textbf{0.9576} & 95.33 & 0.9479 & 95.45 & 0.9485 & 95.85 & 0.9511 \\
\hline
\end{tabular}
\end{table*}

Explainable artificial intelligence, or XAI, is a significant area of AI research and development in an effort to make AI systems more transparent and comprehensible for humans to make decisions. It resolves the black box problem, which often befalls complex DL models like CNNs, by revealing the reasoning behind the findings or forecasts made by AI systems. Additionally, it fosters a sense of accountability and aids users in identifying and reducing biases, errors, and unexpected behaviours in AI systems. XAI uses various techniques and methods, such as feature attribution and visualization, to make AI systems easier to understand and use for experts and non-experts.

GradCAM is an acronym for Gradient-weighted Class Activation Mapping, a computer vision technique. It produces heatmaps that highlight areas of an input image that are most important to a deep neural network's classification decision obtained from the changing gradients during the backpropagation process in DL models. This explanation graphically explains which elements of an image were significant to the model throughout its decision-making process. The heatmaps from GradCAM can directly help visualizations and provide insights into the decision-making processes of DNNs. 

\subsection{Experimental Setup}\label{sec3.5}

The experiments were done in a system with 16GB RAM with 8 cores and an Nvidia RTX3060 GPU with 6GB VRAM. The proposed pipeline to carry out the study done in this work is shown in Figure \ref{fig2}.

Geometric augmentation techniques such as rotation, resizing, zooming, and vertical and horizontal flipping are performed to increase the number of training samples. Transfer learning using pre-trained ImageNet1k weights from the Tensorflow library is done for the EffcientNet backbone to help learn image features quickly. This is followed by the attention mechanisms and the 3-layered MLP classifier with a dropout layer (co-efficient set to 0.1) to prevent overfitting. The softmax activation ensures that the predicted probabilities sum up to one for multi-class classification tasks, making it suitable for classification tasks where each input belongs to exactly one class. Adam is the widely used optimizer in deep learning due to its effectiveness in handling sparse gradients, noisy data, and non-stationary objectives, as it automatically adjusts the learning rate for each parameter, allowing for faster convergence and better optimization of the model's parameters.

The categorical cross entropy (CCE) loss penalizes incorrect predictions more effectively and is used to calculate the classification error during the training backpropagation process (the equation to calculate the loss is given in Equation \ref{eqnL}).

\begin{equation}
\label{eqnL}
\text{Loss}_{\text{cce}} = -\sum_{i=1}^{N} \sum_{j=1}^{K} y_{ij} \log(p_{ij})
\end{equation}

The models were trained for thirty epochs on the augmented dataset, and the efficiency of the proposed architecture is computed with classification metrics such as Top-1 and Top-2 accuracy, precision, recall, and F1-Score were computed on the test data to validate the model's performance. These performance metrics are computed using equations \ref{eqn5} to \ref{eqn8}.

\begin{equation}
\label{eqn5}
\text{Accuracy (Acc)} = \frac{\text{Number of Correct Predictions}}{\text{Total Number of Predictions}}
\end{equation}

\begin{equation}
\label{eqn6}
\text{Precision (Prec)} = \frac{\text{True Positives}}{\text{True Positives} + \text{False Positives}}
\end{equation}

\begin{equation}
\label{eqn7}
\text{Recall (Rec)} = \frac{\text{True Positives}}{\text{True Positives} + \text{False Negatives}}
\end{equation}

\begin{equation}
\label{eqn8}
\text{F1-Score (F1)} = \frac{2 \cdot \text{Precision} \cdot \text{Recall}}{\text{Precision} + \text{Recall}}
\end{equation}

Class-wise metrics from the confusion matrix provide insights into performance for each class. Additionally, the Area Under the Receiver Operating Characteristic (ROC-AUC) curves is evaluated, plotting true positives against false positives to assess the model's performance across different thresholds. Following the availability of the fine-tuned model, the results are also explained visually for clinical correlations using XAI Tools such as GradCAM to gain more insights into the features captured by the model and identify the accurate plane. 

\begin{figure*}[h]
    \centering
    \begin{subfigure}[h]{0.3\textwidth}
        \centering
        \includegraphics[width=\textwidth,height=\textwidth]{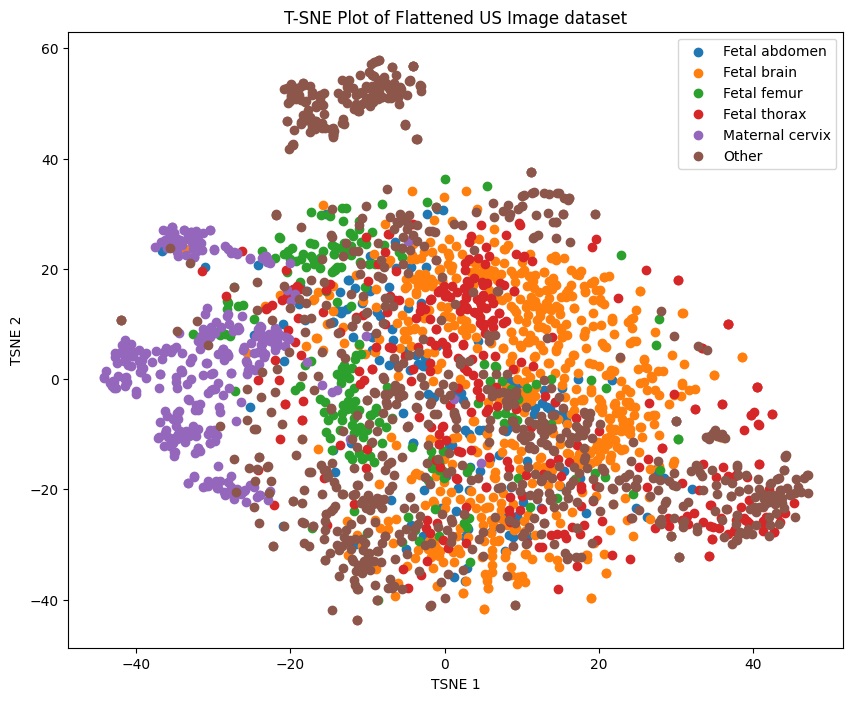} 
        \caption{US Image dataset}
        \label{fig6_1}
    \end{subfigure}
    \hfill
    \begin{subfigure}[h]{0.3\textwidth}
        \centering
        \includegraphics[width=\textwidth,height=\textwidth]{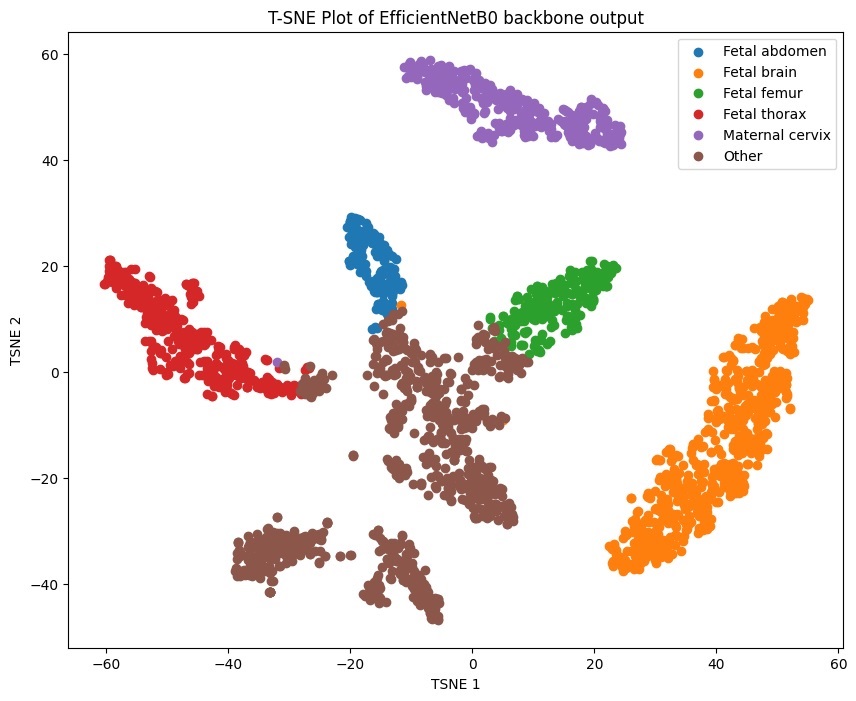} 
        \caption{EfficientNetB0 without MLP}
        \label{fig6_2}
    \end{subfigure}
    \hfill
    \begin{subfigure}[h]{0.3\textwidth}
        \centering
        \includegraphics[width=\textwidth,height=\textwidth]{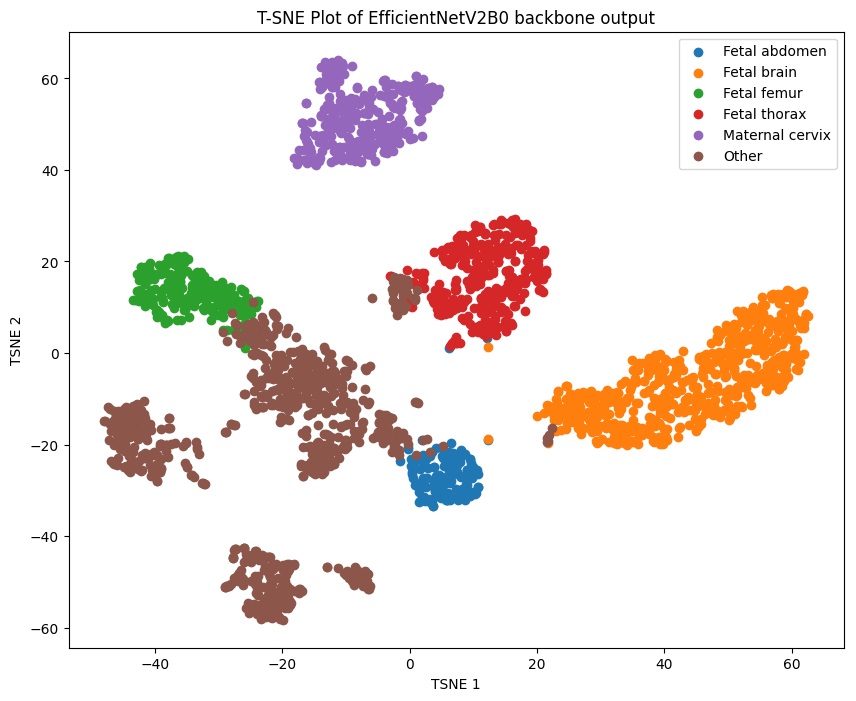} 
        \caption{EfficientNetV2B0 without MLP}
        \label{fig6_3}
    \end{subfigure}
    
    \vspace{0.5cm} % Add some vertical space
    
    \begin{subfigure}[h]{0.3\textwidth}
        \centering
        \includegraphics[width=\textwidth,height=\textwidth]{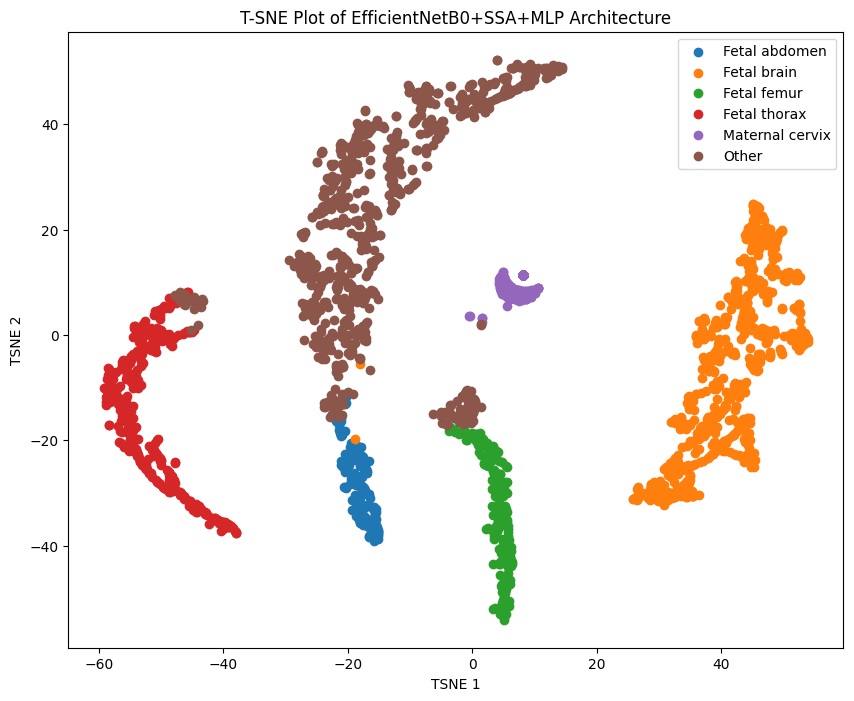} 
        \caption{EfficientNetB0+SSA+MLP}
        \label{fig6_4}
    \end{subfigure}
    \
    \begin{subfigure}[h]{0.3\textwidth}
        \centering
        \includegraphics[width=\textwidth,height=\textwidth]{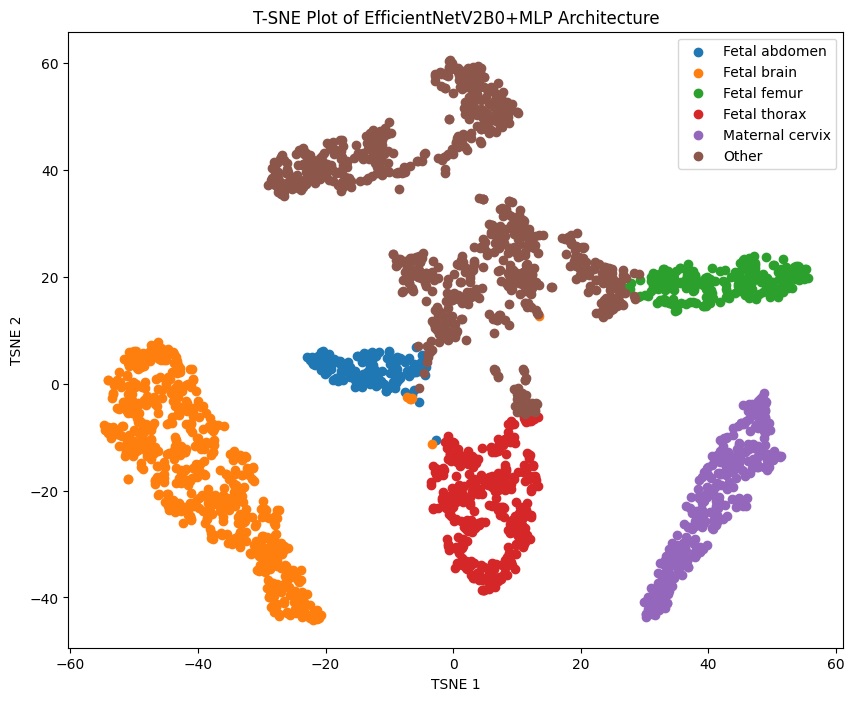} 
        \caption{EfficientNetV2B0+MLP}
        \label{fig6_5}
    \end{subfigure}
    \caption{t-SNE plots of features obtained from the different stages of the proposed architectures: a) Raw Train Dataset Images, b) Features from EfficientNetB0, c) Feature from EfficientNetV2B0, d) Features from EfficientNetB0+SSA+MLP, e) Features from EfficientNetV2B0+MLP}
    \label{fig6}
\end{figure*}

\section{Results and Discussion}\label{sec4}

The authors benchmarked 11 lightweight CNNs from the EfficientNet family, fine-tuning ImageNet1k pre-trained backbones on the FPC dataset. Despite both being EfficientNet variants, EfficientNetBx and EfficientNetV2x differ in design and performance. The metrics of these backbones, with and without the theoretically optimal SSA attention mechanism, are compared in Table \ref{tab3}.

The results show a few generic trends that provide insights into the proposed pipeline's performance. It is also noticed that SSA generally enhances model performance, particularly in Top-1 accuracy and F1 Score, illustrating the benefits of attention mechanisms in capturing dependencies post-feature extraction. For instance, EfficientNetB0's Top-1 Accuracy increased from 95.61\% to 96.05\% and F1 Score from 0.9511 to 0.9534 with SSA. These enhancements indicate that SSA effectively captures dependencies and refines feature representations post-feature extraction, leading to better classification performance. The comparison also highlights EfficientNetV2x models' superior performance, with EfficientNetV2B0 notably offering competitive performance with fewer parameters. As model complexity increases, a trade-off is observed between parameters and performance, with simpler models sometimes outperforming more complex ones due to overfitting in higher-complexity models. EfficientNetV2B0 achieved the highest Top-1 accuracy of 96.25\% and Top-2 accuracy of 99.80\% without an attention mechanism, with only around 6M trainable parameters. This lightweight structure ensures easier deployment in existing healthcare devices. Additionally, EfficientNetB0+SSA achieved a similar Top-1 accuracy of 96.05\% with 1.5x - slightly lower than the former, but with fewer trainable parameters, demonstrating that the attention mechanism enhances model performance by prioritizing crucial feature map values while maintaining much lower parameters.

\begin{figure}[h]
    \centering
    \begin{subfigure}[b]{0.5\textwidth}
        \centering
        \includegraphics[width=\textwidth,height = \textwidth]{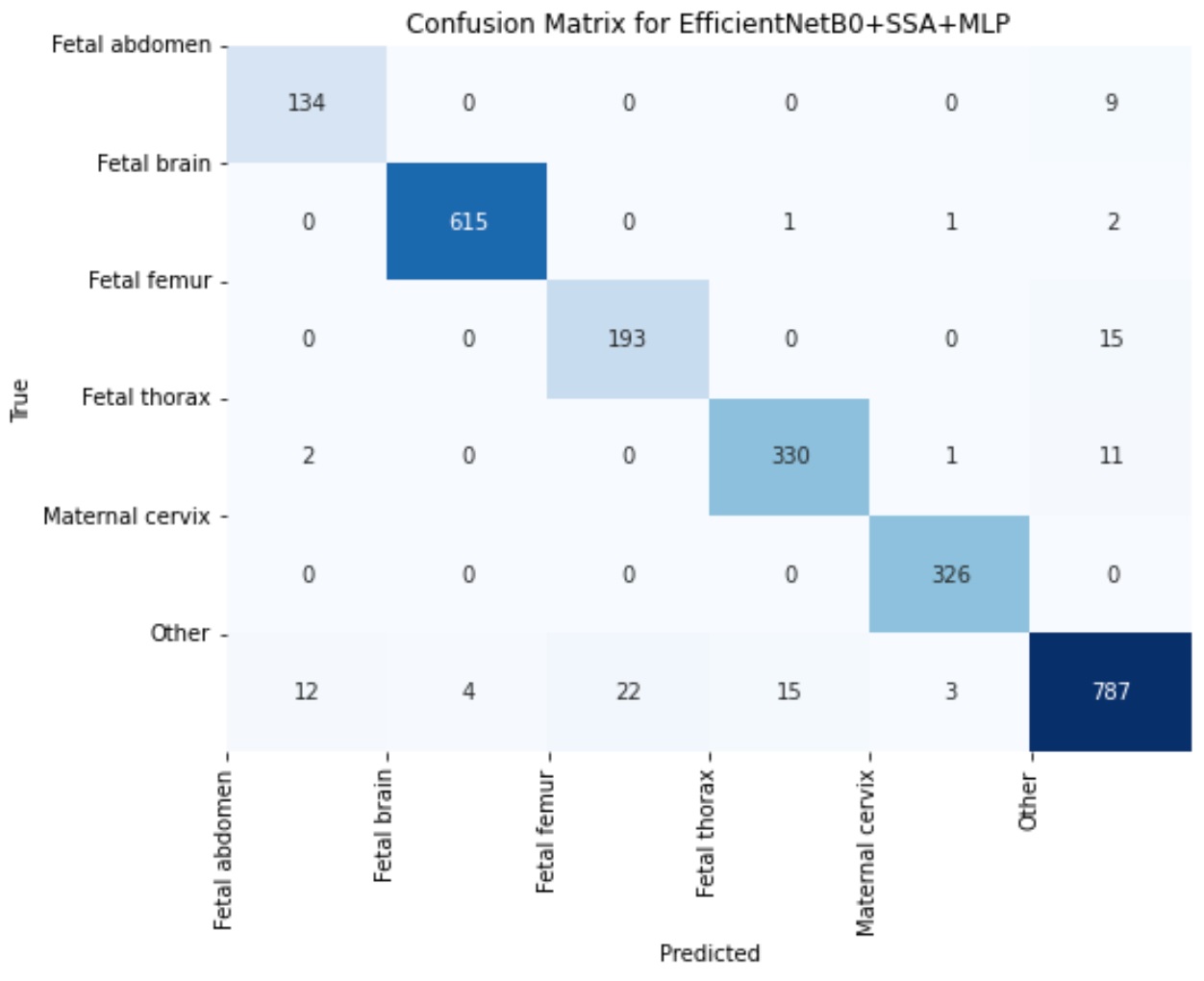} 
        \caption{EfficientNetB0+SSA+MLP}
        \label{fig4_1}
    \end{subfigure}
    \hfill
    \begin{subfigure}[b]{0.5\textwidth}
    \centering
        \includegraphics[width=\textwidth,height = \textwidth]{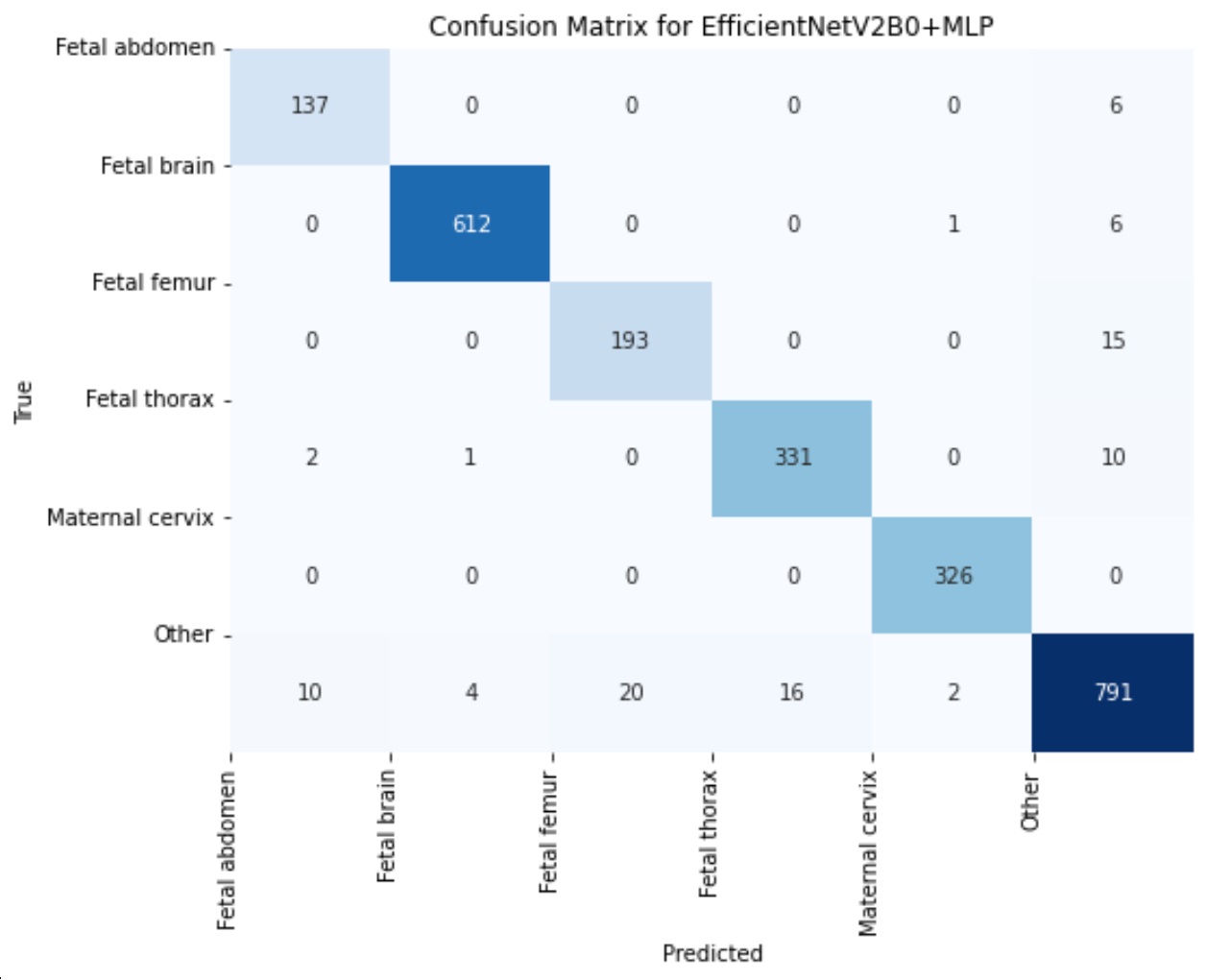}
        \caption{EfficientNetV2B0+MLP}
        \label{fig4_2}
    \end{subfigure}
    \caption{Confusion Matrix of the proposed architectures.}
    \label{fig4}
\end{figure}

\begin{figure}[h]
    \centering
    \begin{subfigure}[b]{0.5\textwidth}
    \centering
    \includegraphics[width=\textwidth,height = 0.75\textwidth]{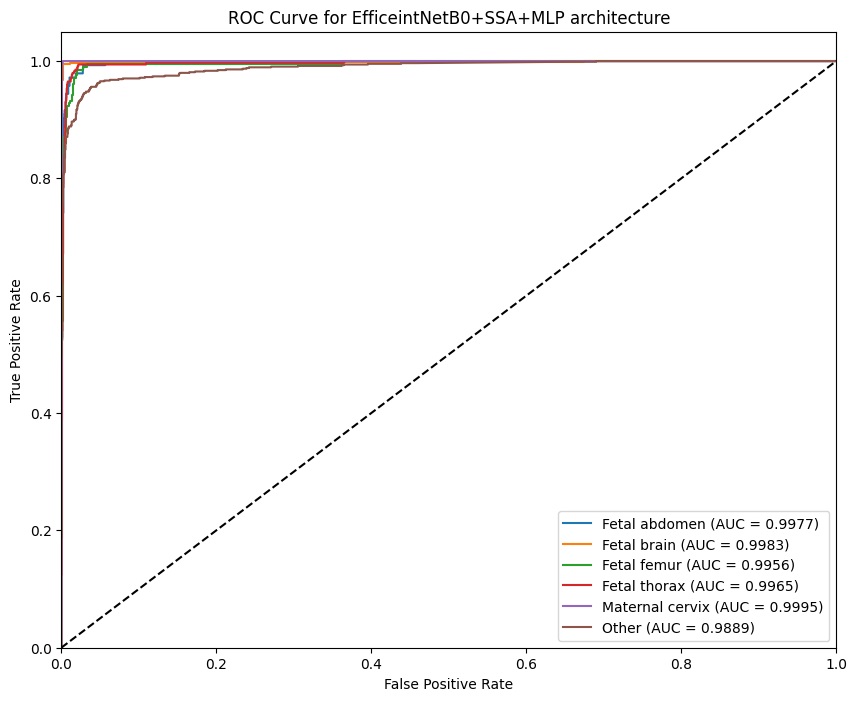}
        \caption{EfficientNetB0+SSA+MLP}
        \label{fig5_1}
    \end{subfigure}
    \hfill
    \begin{subfigure}[b]{0.5\textwidth}
    \centering
    \includegraphics[width=\textwidth,height = 0.75\textwidth]{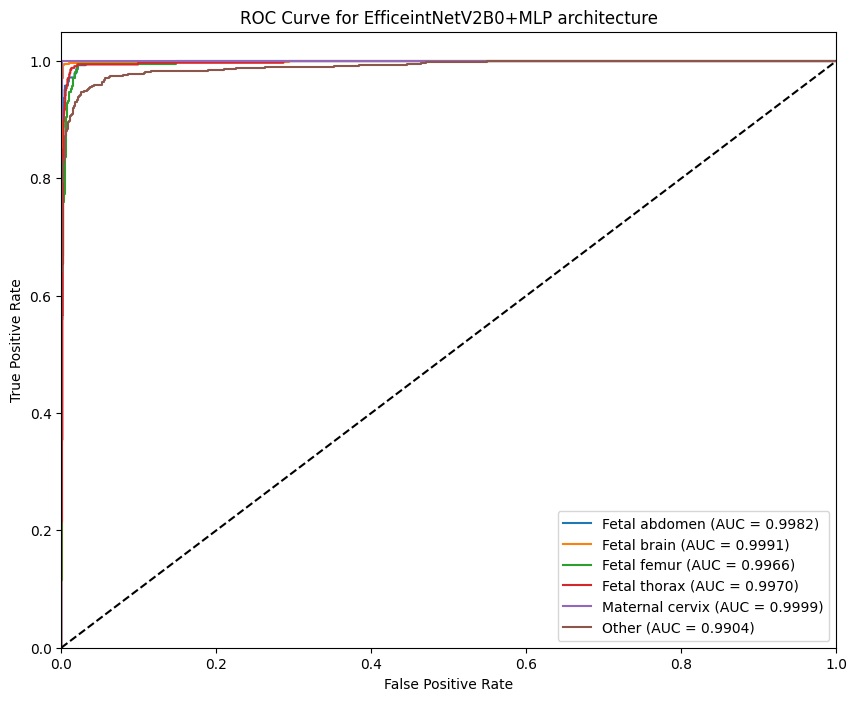} 
        \caption{EfficientNetV2B0+MLP}
        \label{fig5_2}
    \end{subfigure}
    \caption{Receiver Operator Characteristics (ROCs) of the proposed architectures.}
    \label{fig5}
\end{figure}

To evaluate the efficiency of various attention mechanisms and determine potential improvements, experiments were repeated using three different attention mechanisms on the best lightweight backbones (EfficientNetB0 and EfficientNetV2B0). Table \ref{tab4} presents the comparison of metrics across these architectures. SDA shows a mild improvement over no attention, supporting the hypothesis that attention mechanisms refine CNN-extracted features for FPC. MHA, which concatenates multiple SDA heads, performs better than SDA. Nevertheless, as theoretically expected, the EfficientNetB0 with SSA achieves the highest Top-1 accuracy of 96.05\% and an F1-score of 0.9534, demonstrating the effectiveness of self-attention. Interestingly, for EfficientNetV2B0, the backbone without attention surpasses all others, including EfficientNetB0 variants, with the highest Top-1 accuracy of 96.25\% and an F1-score of 0.9576.

\begin{figure*}[h]
    \centering
    \begin{subfigure}[b]{0.32\textwidth}
        \includegraphics[width=\textwidth]{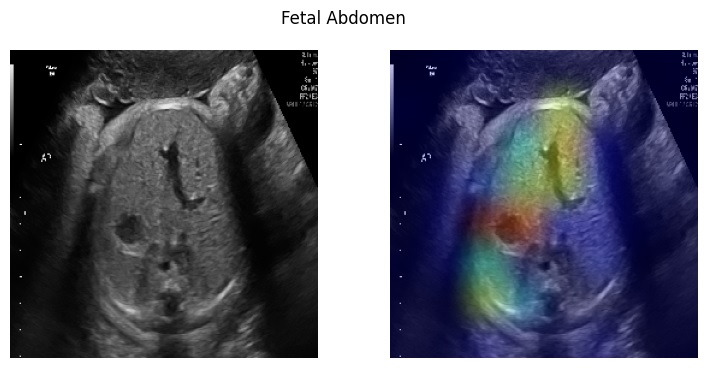} 
     \caption{}
        \label{fig7_1}
    \end{subfigure}
    \hfill
    \begin{subfigure}[b]{0.32\textwidth}
        \includegraphics[width=\textwidth]{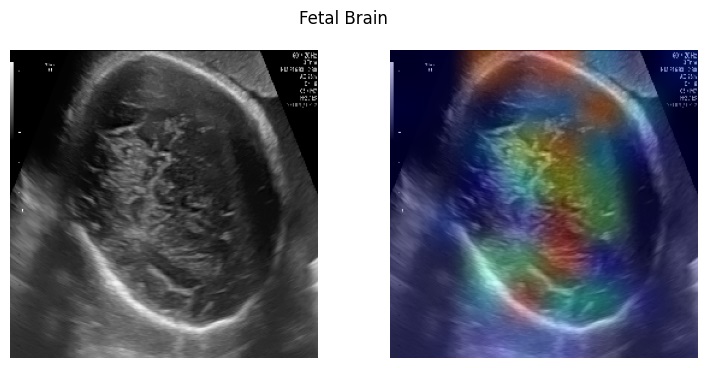} %
        \caption{}
        \label{fig7_2}
    \end{subfigure}
    \hfill
    \begin{subfigure}[b]{0.32\textwidth}
        \includegraphics[width=\textwidth]{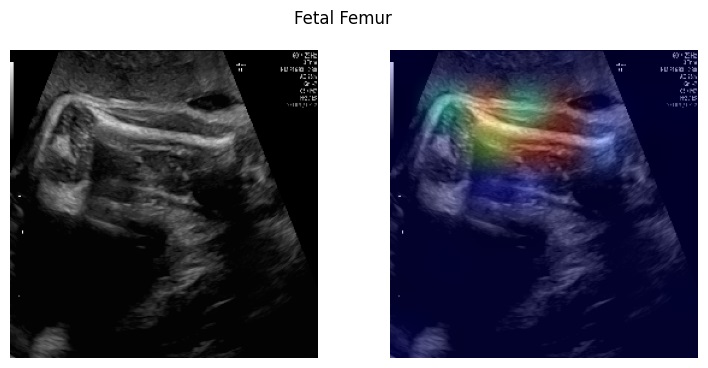} 
        \caption{}
        \label{fig7_3}
    \end{subfigure}
    \hfill
    \begin{subfigure}[b]{0.32\textwidth}
        \includegraphics[width=\textwidth]{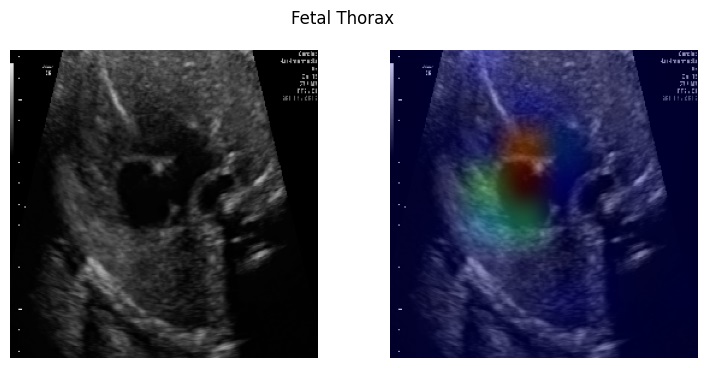} 
        \caption{}
       \label{fig7_4}
    \end{subfigure}
    \hfill
    \begin{subfigure}[b]{0.32\textwidth}
        \includegraphics[width=\textwidth]{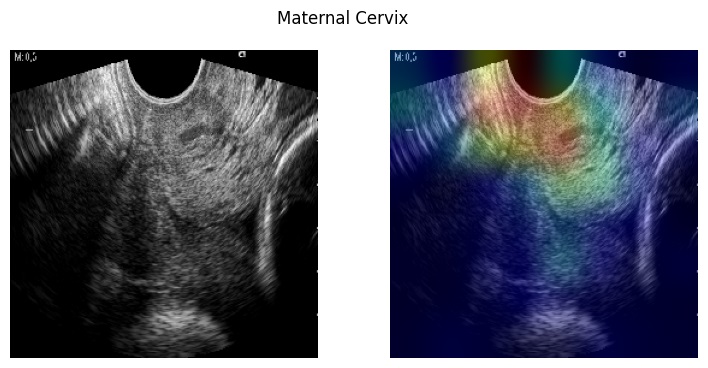} 
        \caption{}
        \label{fig7_5}
    \end{subfigure}
    \hfill
    \begin{subfigure}[b]{0.32\textwidth}
        \includegraphics[width=\textwidth]{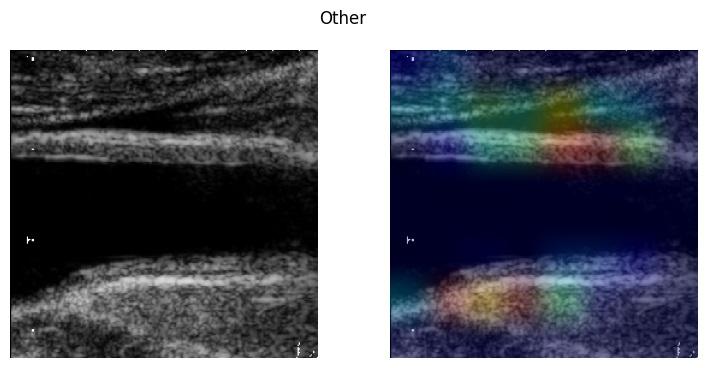} 
        \caption{}
        \label{fig7_6}
    \end{subfigure}
    \caption{A sample of Grad-CAM analysis on correct predictions by the EfficientNetV2B0+MLP architecture.}
    \label{fig7}
\end{figure*}

\begin{figure*}[h]
    \centering
    \begin{subfigure}[b]{0.32\textwidth}
        \includegraphics[width=\textwidth]{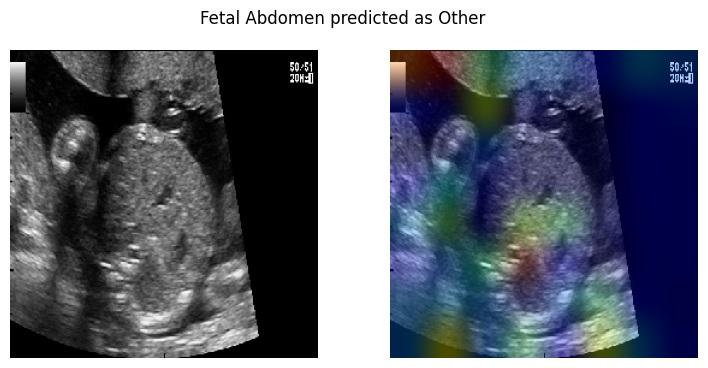} 
     \caption{}
        \label{fig8_1}
    \end{subfigure}
    \hfill
    \begin{subfigure}[b]{0.32\textwidth}
        \includegraphics[width=\textwidth]{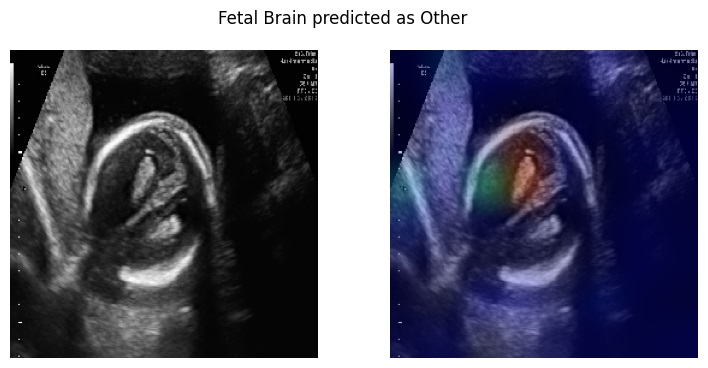} 
        \caption{}
        \label{fig8_2}
    \end{subfigure}
    \hfill
    \begin{subfigure}[b]{0.32\textwidth}
        \includegraphics[width=\textwidth]{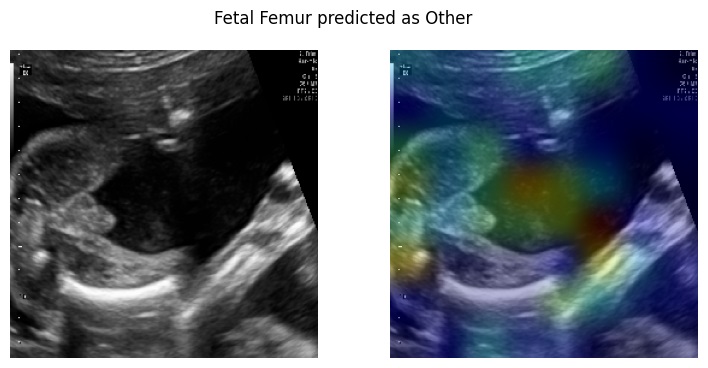} 
        \caption{}
        \label{fig8_3}
    \end{subfigure}
    \
    \begin{subfigure}[b]{0.32\textwidth}
        \includegraphics[width=\textwidth]{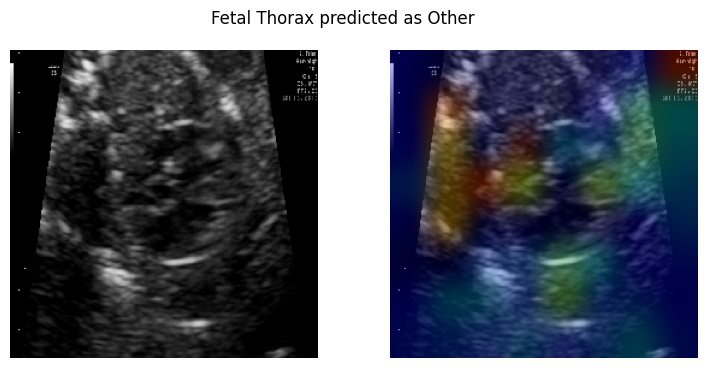}
        \caption{}
       \label{fig8_4}
    \end{subfigure}
    \
    \begin{subfigure}[b]{0.32\textwidth}
        \includegraphics[width=\textwidth]{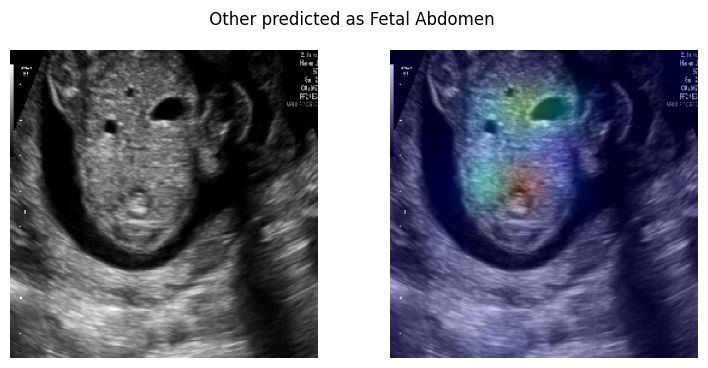} 
        \caption{}
       \label{fig8_5}
    \end{subfigure}
        
    \caption{A sample of Grad-CAM analysis on incorrect predictions by the EfficientNetV2B0+MLP architecture.}
    \label{fig8}
\end{figure*}

To understand this anomaly and determine the importance of MLP in the proposed pipeline, t-distributed stochastic neighbour embedding (t-SNE) is used to map each feature vector from n-dimensional space onto a two or three-dimensional map in order to visualize high-dimensional data. Figure \ref{fig6} shows the 2-D t-SNE plots for various feature representations. In Figure \ref{fig6_1}, the flattened ultrasound image dataset shows data points from all six classes spread across the plot, with same-class points clustering together. Figure \ref{fig6_2} shows EfficientNetB0's classification without MLP fine-tuning, where the fetal brain and maternal cervix are clearly distinguishable, but thorax, abdomen, and femur samples overlap with the 'other' class. In Figure \ref{fig6_3}, EfficientNetV2B0 without MLP shows less overlap, particularly for the 'other' class. Figure \ref{fig6_4} demonstrates that the EfficientNetB0+SSA+MLP architecture accurately classifies the maternal cervix, with embeddings converging at one point. However, Figure \ref{fig6_5} shows that the EfficientNetV2B0+MLP model minimizes inter-class overlaps, creating clearer decision boundaries, thereby also explaining the improved metrics. Thus, t-SNE plots confirm the need for these backbone architectures for the accurate classification of fetal images. They also demonstrate the essence of MLPs for clustering intra-class samples and increasing the distance between inter-class embeddings, enhancing classification accuracy.

To additionally describe the model's classwise performance quantitatively with test samples, Figure \ref{fig4_1} presents the confusion matrix of the EfficientNetB0+SSA+MLP model. The confusion matrix reveals that, apart from the 'other' classes, there are only 5 misclassified samples, leading to a test accuracy of 96.05\%. Notably, there are no false positives for the fetal abdomen, fetal femur, and maternal cervix classes, indicating high specificity for these categories. However, the majority of false positives and false negatives arise from the misclassification of samples in the 'other' class, highlighting the challenge posed by this diverse category. Similarly, Figure \ref{fig4_2} displays the confusion matrix for the EfficientNetV2B0+MLP architecture. This model demonstrates superior performance with only 4 outliers and fewer false positives for the 'other' class samples, resulting in a higher overall accuracy and an F1-Score of 0.9576. The confusion matrices provide a clear view of how well each model distinguishes between different fetal plane classes and where the models struggle, particularly with the 'other' class. 

\begin{table*}[htbp]
\centering
  \caption{Comparison of the performance metrics of different benchmark architectures and the proposed model.}
    \begin{tabular}{|l|c|cccc|}
    \hline
    \textbf{Architecture} & \textbf{Total Parameters} & \textbf{T1-Acc} & \textbf{Prec} & \textbf{Rec} & \textbf{F1} \\
    \hline
    \textbf{CNN-based} &  &  &  & &  \\
    
    VGG-16 \cite{ref43} & 138,357,544 &  92.28 & 0.9078 & 0.9202 & 0.9139 \\
    ResNet101 \cite{ref16} & 44,707,176 & 93.06 & 0.9153 & 0.9359 & 0.9254 \\
    ResNeXt101 \cite{ref44} &  44,707,176 & 93.30 & 0.9187 & 0.9408 & 0.9296 \\
    Inception-v3 \cite{ref45} & 23,851,784 & 93.83 & 0.9197 & 0.9339 & 0.9265 \\
    DenseNet169 \cite{ref48} & 14,307,880 & 93.50 & 0.9251 & 0.9388 & 0.9318 \\
    \hdashline
    \textbf{Transformer-based} &  &  &  &  &  \\
    Vision Transformer \cite{ref46} & 86,567,656 & 93.29 & 0.9213 & 0.9349 & 0.9280 \\
    Swin Transformer \cite{ref47} & 49,606,258 & 93.47 & 0.9351 & 0.9360 & 0.9355 \\
    COMFormer \cite{ref36} & - & 95.64 & 0.9465 & 0.9587 & 0.9523 \\
    \hdashline
    \textbf{Fusion Architectures} &  &  &  &  &  \\
    AlexNet+VGG-16+MLP \cite{ref34} & 159,979,698 & 95.50 & \textbf{0.9552} & 0.9547 & 0.9549 \\
    AlexNet+VGG-19+DarkNet19 \cite{ref37} & 167,374,938 & 95.69 & 0.9402 & \textbf{0.9628} & 0.9508 \\
    \hdashline
    \textbf{Proposed Architectures} &  &  &  &  &  \\
    EfficientNetB0 + SSA + MLP & \textbf{4,133,833} & \textbf{96.05} & 0.9484 & 0.9586 & 0.9534 \\
    EfficientNetV2B0 + MLP & \textbf{6,003,574} & \textbf{96.25} & 0.9529 & 0.9625 & \textbf{0.9576} \\
    \hline
    \end{tabular}
  \label{tab2}
\end{table*}

Figure \ref{fig5} illustrates the ROC curve of the proposed architectures, with AUC values for each fetal ultrasound image class. EfficientNetB0+SSA+MLP achieves the best performance for the maternal cervix, with an AUC of 99.95\%, as further supported by the t-SNE plot in \ref{fig5_1}. This model's lowest performance is in classifying the 'other' classes, with an AUC of 98.89\%, reflecting the earlier confusion matrix observations. In contrast, EfficientNetV2B0+MLP shows strong performance across all classes, achieving an AUC greater than 99\% for all categories. Specifically, it obtains an AUC of 99.99\% for the maternal cervix and 99.82\% for the fetal abdomen, as shown in \ref{fig5_2}. Despite its high overall performance, this model also exhibits lower accuracy in classifying the fetal femur and thorax with 'other' classes, consistent with the confusion matrix results. The ROC curves and AUC values provide a comprehensive understanding of the models' diagnostic abilities, highlighting their high sensitivity and specificity, particularly for the maternal cervix and fetal abdomen classes. These visualizations underscore the models' robustness and the challenges posed by the 'other' class, guiding future improvements in handling diverse and overlapping class features.

The regions of the US fetal images identified by the network models during classification are visualized using Explainable AI (XAI). Grad-CAM heat maps highlight the areas in an input image most influential in determining the classification outcome. This visualization is achieved by computing the gradient of the predicted class score concerning the feature maps in the final convolution layer. The resulting heat map is superimposed on the input image, with red regions indicating the most influential areas and green regions indicating the least emphasized areas. Figure \ref{fig7} is an illustration of some of the correct findings of Grad-CAM analysis performed on the EfficientNetV2B0+MLP. As seen from the heat maps, the gradients are accurately captured based on the contour and the region of the fetus (abdomen, thorax and femur). All samples of the maternal cervix have been perfectly classified, which is also evident from the ROC curves, contributing to the high Top-2 accuracy of the model.

The work has limitations primarily due to natural inter-class variations among fetal planes, resulting in varying classification results for some samples. Figure \ref{fig8} illustrates the GradCam heatmaps for misclassifications observed among test samples. The fetal abdomen (Figure \ref{fig8_1}), brain (Figure \ref{fig8_2}), femur (Figure \ref{fig8_3}), and thorax (Figure \ref{fig8_4}) images are categorized incorrectly into other classes primarily due to low inter-class variance and regional similarities which are indicated by intense red regions of the heatmap. Furthermore, bias in the dataset impacts classification accuracy, particularly within the "other" class, which includes highly variable samples sometimes resembling features of the main fetal planes. This bias leads to more frequent misclassifications, as these samples share characteristics with other fetal plane categories. Addressing these biases in future work, such as by curating cleaner data specifically for the 'other' class, could enhance the model's reliability and accuracy in classifying fetal planes accurately.

Table \ref{tab2} depicts the comparison of the performance metrics of different benchmark architectures and the proposed model. As shown in the table, the architectures which use CNNs alone have the least efficiencies, followed by transformer based architectures. Though Swin transformers perform slightly better than ViT due to sliding kernel attention, the COMFormer architecture could achieve approximately 2\% higher than all metrics over its counterparts due to the cross-covariance transformer. However, it is observed that this study has incorporated two image quality assurance measures to ensure the use of quality US images, altering the dataset before being fed to the network. 

Ensemble bagging methods that combine several backbones have demonstrated better metrics than the single backbone CNN or transformer architectures. For instance, the AlexNet+VGG-16+MLP architecture could correctly classify images as a particular fetal plane out of all images classified as that plane with the highest precision of 0.9552. Though its accuracy is slightly lesser than COMFormers, the F1-Score is higher demonstrating its superiority. Similarly, AlexNet+VGG-19+DarkNet19 architecture effectively captures most instances of a specific fetal plane with the highest recall of 0.9628 and obtains the overall second-best accuracy. However, both these ensemble models have a huge number of trainable parameters for a mild increase in accuracy, which is detrimental in real-time analysis and production deployment scenarios. 

Nevertheless, it is observed that the proposed lightweight CNN+Attention architectures are found to achieve the highest performance metrics compared to the benchmark architectures in the image classification task despite considering all the images in the dataset by incorporating MLP and attention mechanisms to improve feature selection. Though the EfficeintNetV2B0+MLP model doesn't have the best precision and recall scores, it could robustly classify by predicting fewer false positives and false negatives, maintaining a good balance between them to achieve the best F1-score of 0.9576. The metrics are a 1\% improvement over the current results and effectively a 3\% improvement over the metrics of other simple CNN or transformer backbones used in the literature, despite being approximately 40x lighter than them. The EfficientNetB0+SSA+MLP model achieves near-similar metrics while being 1.5x lighter than the former pipeline, suggesting the importance of attention and the scope for future deployments post improvements. 

\section{Conclusion and Future Work}\label{sec5}

Ultrasound imaging has gained wide acceptance among clinicians for monitoring maternal and fetal health due to its non-invasive nature. Integrating AI with ultrasound imaging has the potential to significantly enhance the accuracy, effectiveness, and accessibility of maternal-fetal growth monitoring, ultimately improving the pregnancy process and maternal-fetal health. In this study, a lightweight AI architecture leveraging CNNs and attention mechanisms is proposed to classify the largest ultrasound fetal plane dataset to date, which contains 12,000 images. The model uses EfficientNet backbones (EfficientNet B0 and EfficientNetV2B0) pre-trained on the ImageNet1k dataset, fine-tuned with an attention mechanism and multi-layer perceptrons for classification. The proposed model achieves superior performance with a Top-1 accuracy of 96.25\%, Top-2 accuracy of 99.80\%, and an F1-Score of 95.76, outperforming traditional CNNs and transformer architectures while having 40x fewer trainable parameters. This efficiency facilitates easy deployment on edge devices, enabling real-time FPC to aid clinical practitioners.

The study also incorporates Explainable AI (XAI) techniques using Grad-CAM heatmaps to visualize the regions of ultrasound images identified by the model, aiding in clinical correlation to understand why the model classified an image into a particular fetal plane, aiding in the accurate identification of fetal abnormalities. Performance metrics such as accuracy, precision, recall, F1-score, confusion matrix, ROC curves, and t-SNE plots were used to evaluate the model's effectiveness. Despite its success, the study faced limitations due to low inter-class variance and class imbalance in the dataset, leading to some misinterpretations amongst other planes. Future work will address these limitations by expanding the model to other large ultrasound image datasets and improving feature extraction pipelines for better FPC. The study can also be extended by collaborating with obstetricians and doctors for validation with actual clinical data. Nevertheless, the proposed model's quick deployment capability and interpretability would make it a valuable decision-support tool for monitoring maternal health and fetal abnormalities, enhancing prognosis and treatment plans.

\section*{Author Contributions}
The authors confirm the contribution to the paper is as follows: 

Arrun Sivasubramanian – Conceptualization, Methodology, Software, Writing- Original Draft.

Divya Sasidharan - Software, Writing - Review \& Editing.

Sowmya V – Validation, Writing - Review \& Editing, Supervision.

Vinayakumar Ravi - Writing – Supervision.

All authors reviewed the results and approved the final version of the manuscript.

\section*{Conflict of Interest}
No Conflict of Interest

\section*{Data Availability Statement}
Data will be made available on reasonable request

\section*{Ethical Approval}
None

\section*{Funding statement}
No funding

\section*{Data and Code Availability}
Data will be made available upon reasonable request. The codes will be made available in a GitHub link after the manuscript is published for ease of reproducibility.

\end{document}